\documentclass[12pt]{article} \usepackage{times} \usepackage{geometry}
\usepackage{epsf} \usepackage{caption}
\usepackage{epsfig,wrapfig,paralist}
\usepackage[colorlinks=true,citecolor=blue,urlcolor=blue,linkcolor=blue]{hyperref}
\usepackage{latexsym,graphicx,amssymb,natbib} \usepackage[T1]{fontenc}
\usepackage{hyperref} 
\hypersetup{ colorlinks=true,linkcolor=blue, filecolor=magenta,       urlcolor=cyan, }
 
\usepackage[utf8]{inputenc}
\usepackage{enumitem,amssymb}
\usepackage{ragged2e}
\newlist{thematic}{itemize}{8}
\setlist[thematic]{label=$\square$}
\usepackage{pifont}
\newcommand{\cmark}{\ding{51}}%
\newcommand{\done}{\rlap{$\square$}{\raisebox{2pt}{\large\hspace{1pt}\cmark}}%
\hspace{-2.5pt}}

\newcommand{\lya}{\hbox{Ly$\alpha$}}

\newcommand{\oiii}{\hbox{O\,{\sc iii}}}
\newcommand{\siiii}{\hbox{Si\,{\sc iii}}}
\newcommand{\ciii}{\hbox{C\,{\sc iii}}}
\newcommand{\civ}{\hbox{C\,{\sc iv}}}
\newcommand{\nv}{\hbox{N\,{\sc v}}}

\newcommand{\hii}{\hbox{H\,{\sc ii}}}
\newcommand{\heii}{\hbox{He\,{\sc ii}}}
\newcommand{\llambda}{\lambda\lambda}
\newcommand{\myshrink}{\vspace{-7pt}}
\newcommand{\mylinebreak}{\vspace{8pt}}
\begin{document}
\raggedright
\huge
Astro2020 Science White Paper \linebreak

UV Diagnostics of Galaxies from the Peak of
Star-Formation to the Epoch of Reionization  \linebreak
\normalsize

\noindent \textbf{Thematic Areas:} 
\hspace*{60pt} $\square$ Planetary Systems \hspace*{10pt} $\square$ Star and Planet Formation \hspace*{20pt}\linebreak
$\square$ Formation and Evolution of Compact Objects \hspace*{31pt} $\done$ Cosmology and Fundamental Physics \linebreak
  $\square$  Stars and Stellar Evolution 
\hspace*{1pt} $\square$ Resolved Stellar Populations and their Environments \hspace*{40pt} \linebreak
  $\done$    Galaxy Evolution 
\hspace*{45pt} $\square$             Multi-Messenger Astronomy and Astrophysics \hspace*{65pt} \linebreak

\myshrink
  
\textbf{Principal Author:}

Name:	Casey Papovich 
 \linebreak						
Institution:  Texas A\&M University, College Station, Texas, 77843-4242
 \linebreak
Email: papovich@tamu.edu
 \linebreak
Phone:  979-862-2704
 \linebreak
 
\myshrink
\textbf{Co-authors:}   Dan Stark (University of Arizona),
Steve Finkelstein (University of Texas at Austin),
Swara Ravindranath (STScI),
Danielle Berg (The Ohio State University),
Marusa Bradac (UC Davis),
Mark Dickinson (NOAO),
Ryan Endsley (University of Arizona),
Dawn Erb (University of Wisconsin-Milwaukee),
Nimish Hathi (STScI),
Taylor Hutchison (Texas A\&M University),
Bethan James (STScI),
Intae Jung (University of Texas at Austin),
Jeyhan Kartaltepe (Rochester Institute of Technology),
Anton Koekemoer (STScI),
Ramesh Mainali (University of Arizona),
Sally Oey (University of Michigan),
Naveen Reddy (UC Riverside),
Jane Rigby (NASA Goddard),
Alice Shapley (UCLA),
Charles Steidel (Caltech),
Tommaso Treu (UCLA)
  \linebreak

\myshrink
\textbf{Abstract:}

The rest-frame UV emission from massive stars contains a wealth of
information about the physical nature and conditions of star formation in galaxies.
Using studies of the rest-frame UV, the past decade has witnessed the
beginning of knowledge about the existence and properties of galaxies
during the first few billion years after the Big Bang.   This period
of history corresponds to the formation of the first stars, the rapid
formation of galaxy stellar populations, the reionization of the IGM,
the production and dissemination of heavy elements, and the formation
of the first black holes.  Massive stars in these galaxies
drive all of these events, and their light dominates the  spectral
energy distributions of galaxies.  As we look to the 2020s,
fundamental questions remain about the nature of these stellar
populations and their evolution, from just before the peak of the
cosmic star formation density ($z\sim 3$), up to the epoch of
reionization ($z > 6$).  This next decade will provide transformative
gains both in our ability to identify star-forming galaxies and
accreting supermassive black holes at these early epochs with imaging
surveys in the rest-frame UV  (e.g., LSST, WFIRST).  \textbf{Ground-based,
rest-frame UV spectroscopy on $>$20~m-class telescopes (e.g., GMT/TMT)
offers the ability to investigate the astrophysical conditions in
galaxies at the earliest cosmic times.}  This includes studies of the evolution in galaxy stellar
populations, gas ionization (temperature, pressure), metallicity, and
interstellar (and circumgalactic) gas kinematics and covering
fractions. In this white paper, we
describe the scientific prospects and the requirements for research in
this area. 

\pagebreak




\section{Rest-frame UV Properties of High-Redshift Galaxies}
\myshrink

Measuring the evolution of massive stars and
stellar populations is the key to understanding star-formation and
ionization in the early Universe.   The light from massive stars dominates
all facets that we see in star-forming galaxies including the direct rest-frame UV
continuum ($\sim$1500~\AA) from their photospheres, the nebular
continuum and emission lines from {\sc H~ii} regions they ionize, the
rest optical/near-IR (0.4-2~$\mu$m) emission from their post-main
sequence supergiants, and their far-IR reradiated emission from
dust ($\sim$100~$\mu$m).  Massive stars are the leading
producer of hard, H-ionizing photons, which drive the reionization of
the IGM and produces the last major ``phase change'' in the Universe.
Core-collapse SNe from massive stars are responsible for the
production of nearly all heavy elements (C, O, \ldots, Fe),
particularly in the first few billion years (prior to the onset of
SN~Ia or planetary nebula).  Radiation from massive stars and
energy/momentum from their SNe explosions drive winds, which
distributes metals and sets the kinematics and structure of the
ISM/CGM.  The first massive stars may also produce the first black
holes.  Therefore, to understand the birth and adolescent phases of
galaxies requires that we can describe the
astrophysical properties of their massive stars.  \mylinebreak
 
\begin{figure*}[t]
\begin{center}
\hspace{4mm}
\includegraphics[width=1\textwidth]{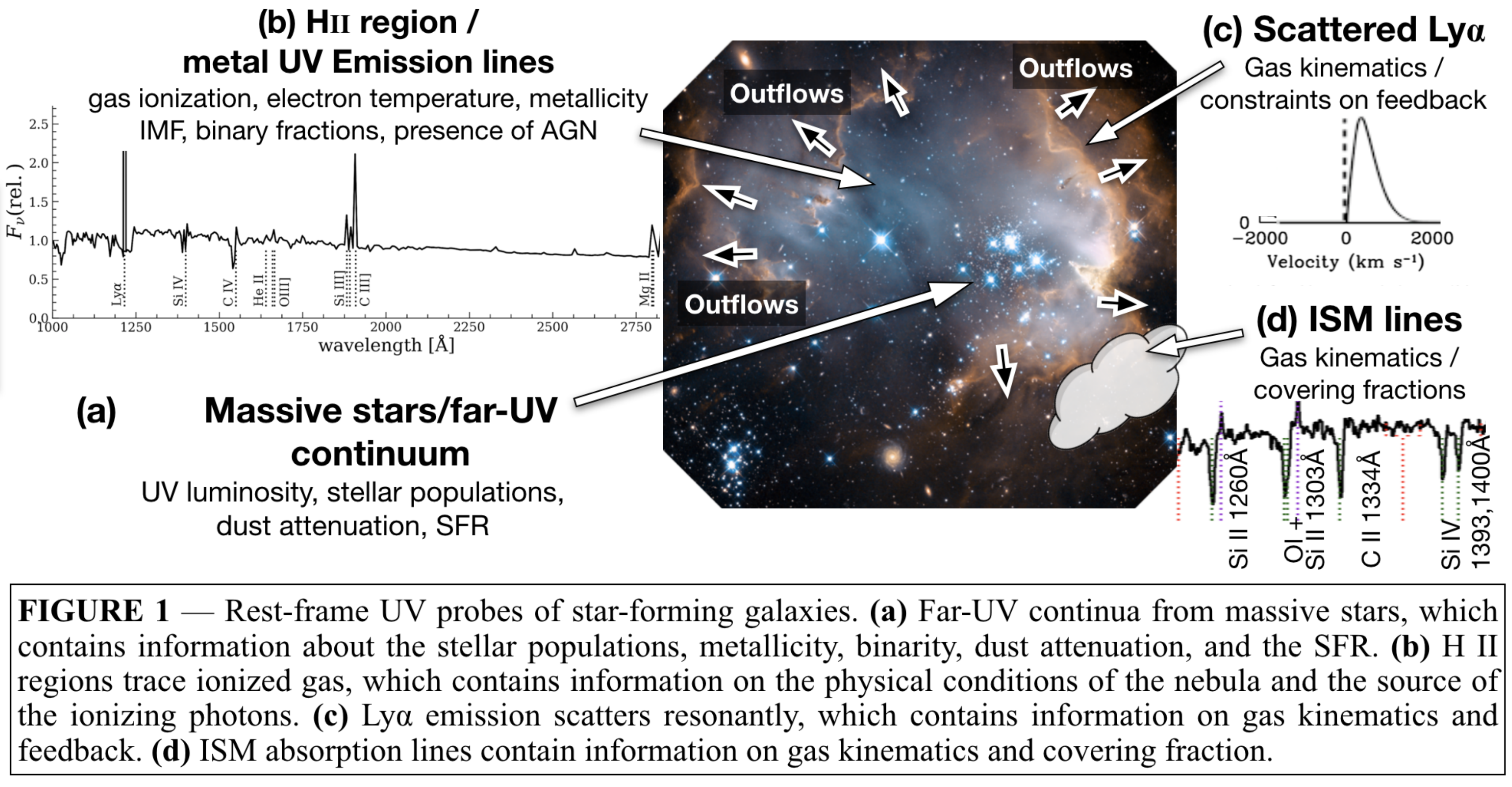}\captionsetup{labelformat=empty}\caption{\label{fig:cartoon}}
  \end{center}
   %
  \vspace{-60pt}
\end{figure*}

The rest-frame UV spectra of star-forming regions contain a rich variety of
information about the massive stars and the physical conditions in
galaxies (Figure~\ref{fig:cartoon}).   These will be accessible to
ground--based telescopes in the 2020s for galaxies from the peak of the cosmic
star-formation rate (SFR) density  ($z\sim 3$) out to the epoch of
reionization (EoR; $z>6$), allowing us to constrain how the evolution
of gas, star-formation, and ionization occurs in galaxies.  \mylinebreak
%
%

The UV spectra contain many emission lines from heavy elements (i.e.,
metals), including  nebular emission from \ciii] $\lambda\lambda
1907$, 1909, \siiii] $\llambda$1883,1892, \oiii]
$\lambda\lambda$1661,1666, He\, {\sc ii} $\lambda$1640, and \civ\
$\lambda\lambda$1548, 1550. These trace the physical conditions
and elemental abundances in the gas, and constrain the source of ionization in these galaxies
\citep{feltre16,gutkin16,jaskot16}.     Nebular UV metal lines also
provide an estimate of the systemic redshift, which when combined with
observations of \lya\ emission and/or absorption features in the
UV constrain the properties of stellar winds, and ISM gas
kinematics/structure and covering fraction
\citep[e.g.,][]{shapley03,jones13,schenker13,erb14,reddy16,sobral18}.
The combination of UV emission lines and UV continuum contain
information about the stellar and gas--phase metallicities, gas density,  the
high-mass end of the IMF, test for the effects of stellar binaries (and
binary fractions), and the origin/nature of the ionizing sources (AGN,
hot stars, X-ray binaries, see, e.g., the BPASS models of
\citealt{eldridge17}; and
\citealt{sommariva12,steidel16,tang18,hutchison19}).    Modeling the
UV continuum and emission lines also  constrain the value of $\xi_\mathrm{ion}$ (the ratio
of the number of ionizing photons ($>$13.6 eV) to non-ionizing
radiation) as this depends sensitively on the properties of the
population of massive stars (age, metallicity, binary fractions, IMF).
Comparing $\xi_\mathrm{ion}$ to direct detections of the escaping
H-ionizing radiation yields measurements of the UV escape fraction
\citep[$f_\mathrm{esc}$; e.g.,][]{steidel18}.  Understanding how
$f_\mathrm{esc}$ correlates with galaxy properties such as \lya\
equivalent width (EW), and properties of the ISM (ionization, covering
fraction, and kinematics) is important as it can be compared to predictions from
simulations to estimate how galaxies contribute to reionization
\citep[e.g.,][]{fletcher18,finkelstein19}.  See also the companion
Astro2020 White Paper by S.~Finkelstein. 
\mylinebreak

\begin{figure*}[t]
\begin{center}
\hspace{4mm}
\includegraphics[trim=3pt 0pt 0pt 0pt, clip=true, width=1\textwidth]{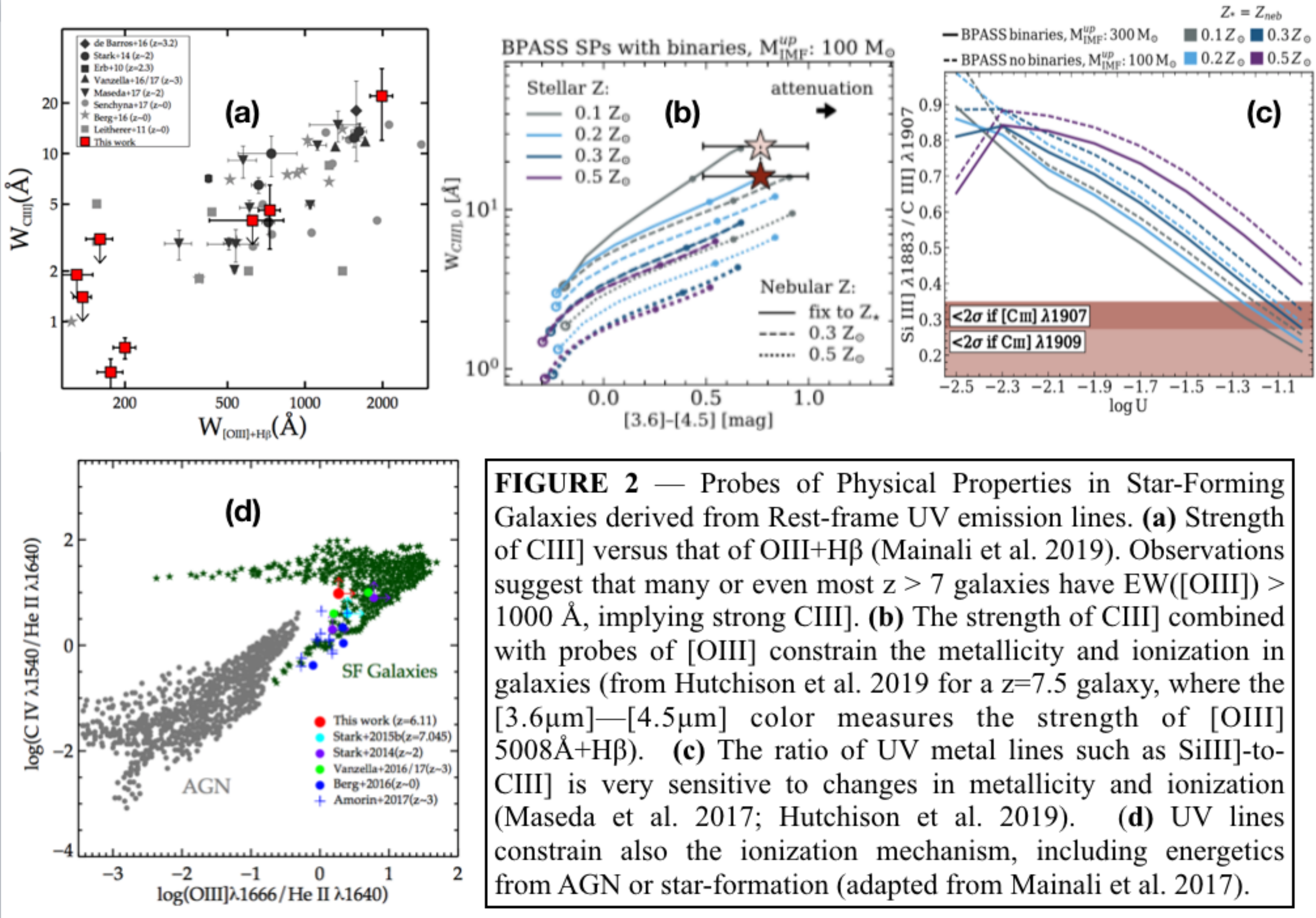}\captionsetup{labelformat=empty}\caption{\label{fig:lineratios}}
  \end{center}
   %
  \vspace{-56pt}
\end{figure*}

One important question to answer in the 2020s is how do the rest-frame
UV properties of galaxies evolve.   Rest-frame UV emission lines from
metals in star-forming galaxies at $z$$\sim$$2$--4 show high
ionization \citep[e.g.,][]{erb10,stark14,steidel16,berg18,mclure18},
with much higher EW compared to  local \hii\ regions
\citep{stark14,shapley15,steidel16,tang18}.    The increase in the
strength of UV emission-lines correlates with the strength of
rest-frame optical lines (such as [\oiii] $\lambda$5007, see
\citealt{stark14,senchyna17}),  which are sensitive probes of
ionization  \citep{kewley15,sanders16,steidel16}.  While rest-frame
optical lines are not at present observable directly at higher redshift
($z$$>$3.8), in some cases the broad-band colors of galaxies (e.g.,
from Spitzer/IRAC, see Figure~\ref{fig:lineratios}) can be used to
constrain the emission line strength 
in the optical \citep[e.g][]{smit15}, which imply very high [\oiii]
EW \citep[e.g.,][]{finkelstein13,roberts-borsani16,stark17,lam19,hutchison19}.
\mylinebreak

Stellar population and photoionization models are now sophisticated
enough that the strength of ratios of UV emission lines can be used to
constrain the nature of the ionizing source, and physical and chemical
properties of
the galaxies \citep{feltre16,gutkin16,jaskot16,stanway16,byler18,nakajima18}.
Currently, it is very difficult to compare these predictions with
observations.   One problem is the UV spectral features at $z$$\sim$3
are weak:  the average Lyman-Break Galaxy (LBG) shows rest-frame EW
1.7, 1.3, and 0.2~\AA\ for \ciii], \heii, and \oiii], respectively
(\citealt{shapley03};  see below and Figure~\ref{fig:lineratios}).   A
more serious problem is that the distant galaxies are very faint
($m$$>$26~AB mag), accessible only at the very limit of 8-10~m
telescopes.   As a result, observations of UV metal lines exist for
only  a handful of the brightest high-redshift galaxies  ($<$20 galaxies at
$z$$>$5, including only 5 galaxies with $z$$>$7, where detections are
so rare each one typically produces a separate publication,
\citealt{stark15,stark15b,stark17,mainali18,hutchison19}).
\mylinebreak

Surprisingly, the intensity of the UV metal emission lines in  $z$$>$5
galaxies is an order of magnitude larger than was predicted from
observations of galaxies at lower redshifts, suggesting that the
nature of reionization-era galaxies may be very different than at
$z$$\lesssim$3
\citep[e.g.,][]{stark16,stark17,ding17,maseda17,matthee17,mainali18,hutchison19}. The
ionization in these star-forming regions is extremely high ($\log
U$$>$$-2$, where $U$ is the ratio of the number density of ionizing
photons to Hydrogen atoms).  In rare cases \civ\ and \nv\ emission is
seen \citep{tilvi16,hu17,laporte17,mainali17}, indicating extreme
ionization in interstellar metals,
requiring an even more energetic supply of radiation (i.e., very young
hot stars or AGN, Figure~\ref{fig:lineratios}).
%
%

\myshrink
\section{UV Spectroscopy of High-Redshift Galaxies in the 2020s} 
\myshrink

\begin{figure*}[t]
\begin{center}
\hspace{4mm}
\includegraphics[width=1.0\textwidth]{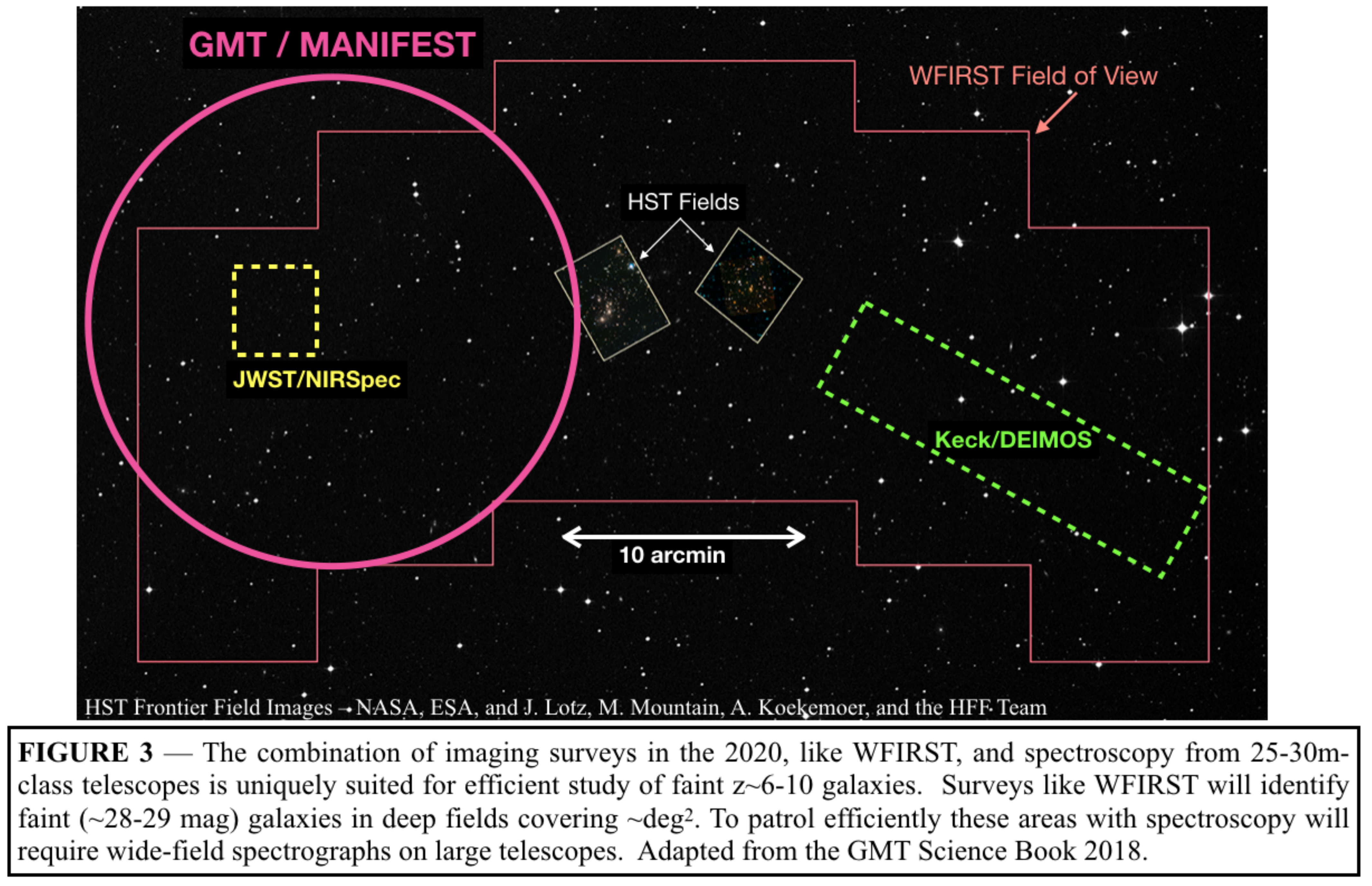}\vspace{-6pt}\captionsetup{labelformat=empty}\caption{
\label{fig:wfirst}
}
  \end{center}
  \vspace{-24pt}
   %
  \vspace{-24pt}
\end{figure*}

The discovery space for science derived from UV diagnostics
is huge, but requires spectroscopy sensitive to galaxies currently
beyond the limits of our telescopes.  
%
%
%
By the mid-2020s, imaging surveys from facilities such as WFIRST and
LSST will start to deliver large samples of high-redshift LBGs
(3$<$$z$$<$10) selected over wide areas and faint magnitudes
(including $\sim$2,000 deg$^2$ fields to $m_\mathrm{AB}$=26.5,  and
$\sim$1 deg$^2$ fields to $m_\mathrm{AB}$=28--29).     Large
telescopes will be required to obtain the rest-frame UV spectroscopy
of these galaxies.  Starting in $\sim$2021, JWST will be very
efficient at measuring spectra of early star forming galaxies, but its
field of view is small and lifetime short (Figure~\ref{fig:wfirst}).
To capitalize on the depth/area from imaging surveys of the 2020s and
beyond, we will require multiplexed, wide-field spectroscopy from
$>$20~m telescopes (e.g., the GMT and TMT; see
Figure~\ref{fig:wfirst}, where the GMT with the MANIFEST fiber
positioner system will  enable deep spectroscopic surveys of $\sim$1
deg$^2$ fields).   Spectroscopy from $>$20~m telescopes will operate
contemporaneously with LSST and WFIRST surveys, and will provide
greater sensitivity and higher resolution at near-IR wavelengths
($\sim$1--2~$\mu$m) compared to JWST.  This will be
paramount for studying the rest-UV properties of newly discovered
galaxies at 5$<$$z$$<$9.\footnote{See:
\url{https://jwst.stsci.edu/science-planning/proposal-planning-toolbox/jwst-sensitivity-and-saturation-limits}
and \url{https://www.gmto.org/sciencebook2018/}} \mylinebreak

\begin{figure*}[t]
\begin{center}
\hspace{4mm}
\includegraphics[width=0.95\textwidth]{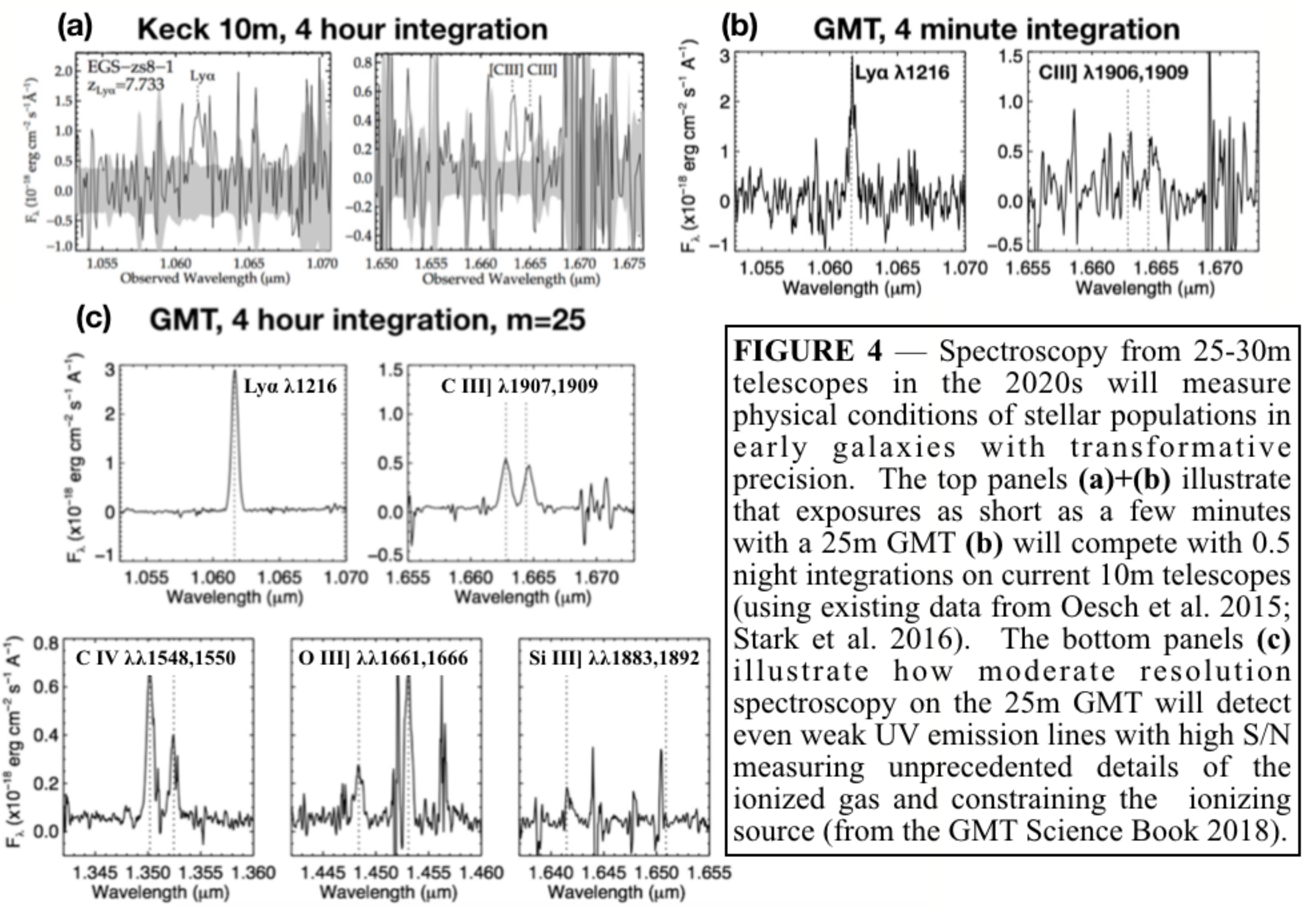}\captionsetup{labelformat=empty}
\caption{
%
%
\label{fig:eltlines}}
  \end{center}
   %
  \vspace{-58pt}
\end{figure*}

Optical and near-IR spectrographs on 25-30~m telescopes will enable
the mainstream study of the rest frame UV emission in high--redshift
galaxies. The efficiency gained by spectroscopy with 25-30~m--class
telescopes is illustrated in Figure~\ref{fig:eltlines}, which allows
high S/N emission-line studies of $z$$\sim$6$-$10 galaxies.  The
majority of these galaxies will be very faint ($m_\mathrm{AB}$=27--28)
detected from surveys operating in the 2020s (such as LSST and
WFIRST).    A host of fainter emission lines (e.g., \civ, \oiii],
\siiii], \ciii]) will become apparent in moderate exposures
($\sim$1--4 hrs), enabling the characterization of their ionizing
source (hot stars and/or AGN), metal content, and gas properties 
\citep[e.g.,][]{mainali17,hutchison19}.  Longer exposures
($\sim$10--30 hrs) on $25-30$~m telescopes would enable emission line
studies to be extended to even fainter ($m_\mathrm{AB}$$\sim$28--29)
galaxies, providing insight into how the physical characteristics of
ionization and metal fraction depend on galaxy luminosity/mass, and
informing us about the properties of the more representative
population of low mass galaxies. Spectroscopy with 25--30m telescopes
of fainter galaxies (or gravitationally lensed systems) will enable
studies to be extended to the very highest redshifts ($z>10$), probing
the earliest phases of galaxy formation. \mylinebreak

Wide-field areal coverage combined with high multiplexing and the
large apertures of 25-30m telescopes provides enormous efficiency
gains in the study of faint ($z$$\sim$6--10) objects, where  current
8–10 m class telescopes are not sensitive enough to detect faint
emission lines from $m_\mathrm{AB}$=26--29 galaxies.    High multiplexing
over large fields-of-view is also required as these galaxies have
surface densities of only $\sim$1 per arcmin$^2$ at $m_\mathrm{AB}$=27
\citep{finkelstein16}.  The survey speed provided by the
25--30~m-class telescopes is a huge advantage for this science,
compared to JWST or smaller ground-based telescopes
\citep[see][]{newman15}. 

%
%
%
%
%
%
%

\myshrink
\section{The Search for ``Metal-free'' Epochs of Star-Formation} 
\myshrink

The telescope facilities of the 2020s could discover the first known
galaxies with “metal-free” stars.  Such galaxies would show strong
emission from hydrogen and helium (i.e., \heii\ $\lambda$1640) with no
UV lines from other heavier elements. This discovery would be
transformative in that it would provide the first glimpse of how
metal-free stars behave.  While galaxy formation is well underway by
the end of reionization, the metal enrichment provided by star
formation is a local and gradual process. Theoretical
investigations predict that Pop III stars can still form in chemically
pristine gas in large, under-dense regions at $z$$\sim$6, one billion years after the Big
Bang \citep[e.g.,][]{scannapieco03,trenti09,xu16b}. 
%
%
If the Pop III IMF
is weighted toward very high mass stars \citep[e.g.,][]{bromm04}, the strong UV radiation field
will power strong \heii\ $\lambda$1640 emission \citep[e.g.,][]{schaerer02}.  \mylinebreak

Wide-field, multiplexed optical and near-IR spectrographs on 25-30~m
telescopes will be able to characterize the strength of \heii\ and
metal lines in thousands of galaxies at redshifts $z$$>$6 over
1~deg$^2$ fields,  thereby enabling the discovery of Pop III stars.
Because the Pop III phase is expected to be very short ($\sim$10 Myr),
and may exist only in under-dense regions,  these objects are very
rare.   The \lya\ luminosity from Pop III galaxies is expected to be
$\lesssim$$10^{43}$~erg s$^{-1}$ \citep{scannapieco03}, with the
\heii\ $\lambda$1640 emission being $\sim$0.1$\times$$L(\lya)$ and no
other metal lines present \citep{schaerer02}.     To detect \heii\ would require only 1-2~hr
exposures on 25-30~m class telescopes, allowing the efficient
discovery of these rare sources over large fields. 

\pagebreak
\bibliographystyle{aasjournal}
\bibliography{astro2020}

\end{document}